\documentclass[]{revtex4}
\usepackage{graphicx}
\usepackage{amssymb}
\usepackage{bm}

\begin{document}

\title{Immersive VR Visualizations by VFIVE. Part 2: Applications}

\author{Akira Kageyama}
\address{Graduate School of System Informatics, Kobe University, \\
Kobe, 657-8501, Japan\\
kage@cs-kobe-u.ac.jp}

\author{Nobuaki Ohno}
\address{University of Hyogo, \\
Kobe, 650-0047, Japan}

\author{Shintaro Kawahara}
\address{Earth Simulator Center, Japan Agency for Marine-Earth Science and Technology, \\
Yokohama, 236-0001, Japan}

\author{Kazuo Kashiyama}
\address{Department of Civil and Environmental Engineering, Chuo University, \\
Bunkyo-ku, 112-8551, Japan}

\author{Hiroaki Ohtani}
\address{National Institute for Fusion Science, Toki, 509-5292, Japan\\
The Graduate University for Advanced Studies (SOKENDAI), Toki, 509-5292, Japan}

\begin{abstract}
VFIVE is a scientific visualization application
for CAVE-type immersive virtual reality systems.
The source codes are freely available.
VFIVE is used as a research tool in various VR systems.
It also lays the groundwork for developments
of new visualization software for CAVEs.
In this paper, we pick up five CAVE systems
in four different institutions in Japan.
Applications of VFIVE in each CAVE system are summarized.
Special emphases will be placed 
on scientific and technical achievements made possible by VFIVE.
\end{abstract}

\keywords{virtual reality, CAVE system, immersive display system}

\maketitle

\section{Introduction\label{sec:intro}}

VFIVE is a scientific visualization application
for CAVE-type immersive virtual reality (VR) systems.
Visualization methods and user-interface implemented
in this application are described in detail 
in our complemented paper\cite{Kageyamaa}.

We started developing visualization software for a 
CAVE in late 1990s, to visualize data of our own simulations\cite{kageyama1998}.
The applications gradually attracted interests 
of other simulation researchers, and getting their feedbacks, 
we developed a general-purpose visualization application
VFIVE\cite{kageyama1999,Kageyama2000}.

We have been improving VFIVE for more than a decade. 
It is brought up to be a practical tool 
for three-dimensional, interactive visualization in CAVEs
through the continued improvements.

VFIVE is coded with C++ language with OpenGL and CAVElib.
Its source codes are freely distributed.
The purpose of this paper is to summarize its
applications and, especially, achievements made possible by VFIVE,
in several CAVE systems in Japan.

%-----------------------------------------------------------------------------------
\section{$\pi$-CAVE at Kobe Univ.\label{sec:picave}}
%-----------------------------------------------------------------------------------

%-----------------------------------------------------------------------------------
\subsection{Hardware}
%-----------------------------------------------------------------------------------

\begin{figure}[h]
  \begin{center}
%=================<fig>=================
   \includegraphics[width=0.40\linewidth]{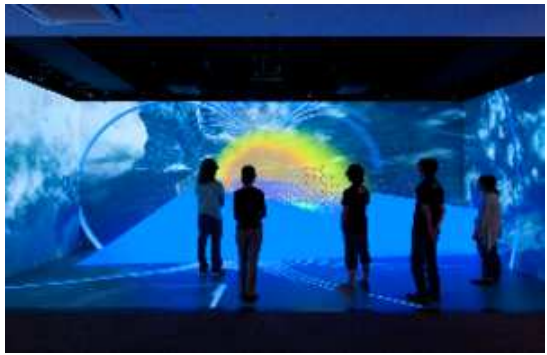}
  \caption{%
    Overview of $\pi$-CAVE at Kobe Univ.
    A rectangular parallelepiped CAVE system with side lengths 3m $\times$ 3m $\times$ 7.8m.
  }
  \label{fig:picave-overview}
%=================</fig>=================
  \end{center}
\end{figure}

$\pi$-CAVE is a CAVE system installed in 2011  
at Integrated Research Center (IRC), Kobe University.
It is named after IRC's location, Port Island (PI).
An overview of the $\pi$-CAVE system is shown in Fig.~\ref{fig:picave-overview}.

$\pi$-CAVE has a rectangular parallelepiped configuration with side 
lengths of 3m $\times$ 3m $\times$ 7.8m. 
To the authors' knowledge, this is the largest CAVE system in Japan.
Its wide width (7.8m) is one of the characteristic features of this CAVE system. 
It enables several people to stand on the floor at the same time, 
without causing mutual occlusion of the screen views in the room.

\begin{figure}[h]
  \begin{center}
%=================<fig>=================
   \includegraphics[width=0.5\linewidth]{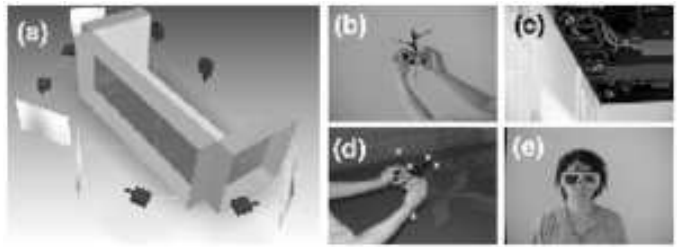}
  \caption{%
    Overview of $\pi$-CAVE at Kobe Univ.
    A box-shaped CAVE system with side lengths 3m $\times$ 3m $\times$ 7.8m.
  }
  \label{fig:picave-projector-trackers2}
%=================</fig>=================
  \end{center}
\end{figure}

$\pi$-CAVE has four screens; three wall screens (front, right, and left) and a floor. 
Semi-transparent soft screens are used for the walls. 
The images on them are rear-projected. 
Six sets of Digital Light Processing (DLP) projectors are used;
see Fig.~\ref{fig:picave-projector-trackers2}a.
The floor is a hard screen on which stereo image is projected from the ceiling. 
Two projectors are used to generate front wall images. 
The edge blending technique is applied on the interface between the projections.
Another pair of projectors is used for the floor screen. 
The resolution of the DLP projector (Christie WU12K-M) 
is 1920 $\times$ 1200 pixels. 
The brightness is 10,500 lumens.

An optical motion tracking system (Vicon) is used for the head and the wand tracking 
(Fig.~\ref{fig:picave-projector-trackers2}b to e).
Ten cameras with 640 $\times$ 480 resolution are installed on the top of the wall screens. 
A commonly used commercial API (Trackd) is used for the interface.

Two computer systems are used for the computation and the rendering.
One is a Linux-PC (HP Z800) with 192GB shared memory. 
Three sets of external GPU (NVIDIA QuadroPLEX) are 
used for the real time and stereoscopic image generation for the six projectors. 
Another computer system is a Windows-PC cluster system. 
VFIVE runs on both computer systems.

%-----------------------------------------------------------------------------------
\subsection{VFIVE Applications}
%-----------------------------------------------------------------------------------

One of the most actively applied fields 
of VFIVE in $\pi$-CAVE is visualization of 
MHD (magnetohydrodynamics) simulations.
Since an MHD simulation produces
multiple vector fields such as magnetic field and flow field, 
grasping their mutual interaction is a demanding task for 
usual visualizations on PC monitors.
VFIVE's visualization methods for vector fields play important roles in their analysis.

\begin{figure}[h]
  \begin{center}
%=================<fig>=================
   \includegraphics[width=0.5\linewidth]{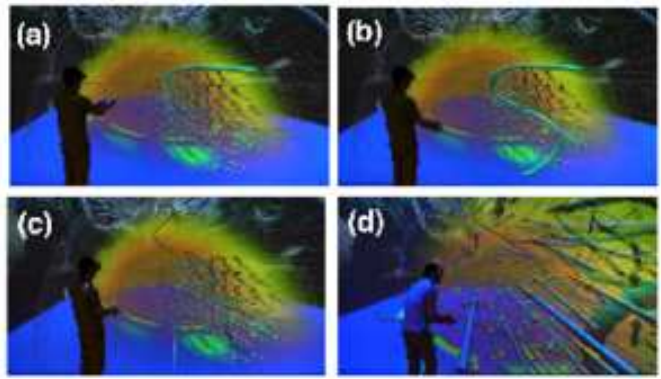}
  \caption{%
    A new visualization component implemented in VFIVE at $\pi$-CAVE system.
    This method shows the stretch-twist-fold process of a magnetic field
    lines in a magnetohydrodynamic fluid simulation.
  }
  \label{fig:tube_advector_combined}
%=================</fig>=================
  \end{center}
\end{figure}

New visualization components are also developed at $\pi$-CAVE,
on the groundwork of VFIVE.
An example that has recently been incorporated into VFIVE is 
``Tube Advector''.

Magnetic field lines in an MHD fluid such as
liquid iron of the Earth's outer core or
hydrogen plasma in the solar convection zone
can be regarded as if they are ``frozen'' into the fluid flow
when the electrical resistivity is negligibly small\cite{Davidson2001}.
This frozen-in nature of magnetic field lines in MHD systems
plays an important role in the magnetic field generation process in celestial bodies.

A magnetic field generation can be regarded as an
increase process of field line density going through a unit area.
However, no magnetic field line comes from nothing,
because the magnetic field is a divergence-free vector field:
A magnetic field line never has an open end.
To increase the number density of magnetic field lines,
a field line has to be stretched, twisted, and folded.

To visualize the stretch-twist-fold process 
of frozen-in magnetic field lines, 
a new VFIVE visualization method, ``Tube Advector'', was developed\cite{Murata2011a}.
Sequential snapshots of a VFIVE visualization of geodynamo simulation data
using this ``Tube Advector'' 
are shown in Fig.~\ref{fig:tube_advector_combined}.
In this figure, the user specifies
an S-shaped curve of a magnetic field line
with a short beam emitted from the portable controller wand
(Fig.~\ref{fig:tube_advector_combined}a and b).
Reflecting the frozen-in nature, this field line is conveyed, or advected, by the flow~(c).
The user can analyze the thee-dimensional advection motion,
observing its stretch-twist-fold process in detail~(d).

Recently in Kobe University, a middleware called Multiverse for CAVE systems 
was developed\cite{Kageyama2011b,Kageyama2011}.
Multiverse is a kind of application launcher or a ``3-D Desktop'' in the VR space.
Multiple CAVE applications can be loaded into Multiverse and 
3-D icons with virtual touch screens depict them.
By touching one of the icons floating in the air with the wand,
the corresponding application is launched.
VFIVE is loaded into Multiverse in $\pi$-CAVE.
The VFIVE application
shown in Fig.~\ref{fig:picave-projector-trackers2} is actually launched
from the Multiverse environment.

%-----------------------------------------------------------------------------------
\section{pCAVE at Kobe Univ.}
%-----------------------------------------------------------------------------------

%-----------------------------------------------------------------------------------
\subsection{Hardware}
\begin{figure}[h]
  \begin{center}
%=================<fig>=================
   \includegraphics[width=0.25\linewidth]{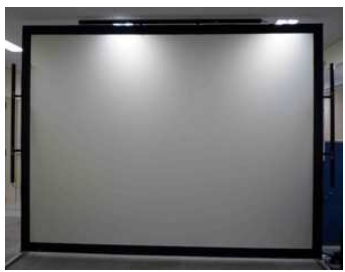}
  \caption{%
    pCAVE system at Kobe University.
    One screen VR system with the head and hand tracking system.
  }
  \label{fig:120124_pCAVE_screen}
%=================</fig>=================
  \end{center}
\end{figure}

%-----------------------------------------------------------------------------------

pCAVE, which is shown in Fig.~\ref{fig:120124_pCAVE_screen},
is a one-screen VR system at Kobe Univ.~with a rear stereo projector.
The screen size is W3260mm $\times$ H2500mm. 
A DLP projector (Christie Mirage S+4K) is used.
The computer system is SGI Asterism Deskside ADT08C with 
Quadcore AMD Opteron 2350 and 4GB memory.
NVIDIA Quadro FX4600 is used for the GPU.

The head and hand tracking system used in pCAVE is 
Intersense IS-900, which is an 
inertial and ultrasonic hybrid system.
CAVELib (ver.~3.2) and VR Juggler\cite{Meno2012} are used as basic APIs for 
applications on pCAVE.
The current version of VFIVE on pCAVE uses only CAVElib.

%-----------------------------------------------------------------------------------
\subsection{VFIVE Applications}
%-----------------------------------------------------------------------------------

pCAVE system is mainly used as a workbench for
CAVE application developers at Kobe University,
before porting to $\pi$-CAVE system described in section~\ref{sec:picave}.
Development of a new visualization method,
that is to be incorporated into VFIVE in future,
is first developed and tested at pCAVE.

\begin{figure}[h]
  \begin{center}
%=================<fig>=================
   \includegraphics[width=0.5\linewidth]{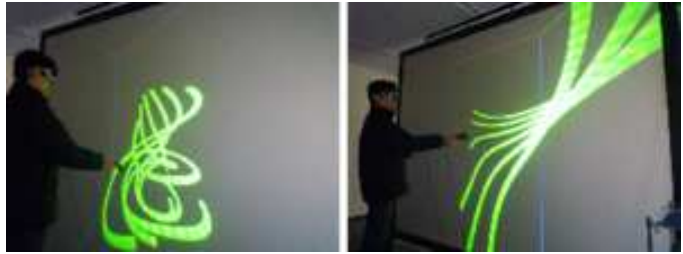}
  \caption{%
    A new visualization method of VFIVE under development at pCAVE system.
　This is a visualization method for vector fields.
    Multiple field lines are traced from a short virtual beam emitted from the wand.
  }
  \label{fig:yoshizaki-realtime-flow-lines}
%=================</fig>=================
  \end{center}
\end{figure}

An example of new VFIVE visualization
that is under development at pCAVE
is ``Interactive Field Lines'', shown in
Fig.~\ref{fig:yoshizaki-realtime-flow-lines}.
In ``Particle Tracer'' method or ``Field Line'' method
implemented in the original VFIVE,
the seed points of the line tracings
are fixed in space after the user's specification by the wand button pressing.
In contrast to them,
the seed points of the ``Interactive Field Lines''
move in real time, following the user's hand (or wand) motion.
After selecting this visualization method from VFIVE's menu,
a short beam appears from the wand's tip.
Several seed points are located on the beam.
As the user moves the wand,
field line integration with a 4-th order Runge-Kutta method
of selected vector field is performed in real time.
The user can intuitively grasp (possibly) complicated 
three-dimensional structure of the vector field,
by moving his/her hand and observing not only the instantaneous shape 
of the field lines, but also their transformation process.
Each field line in this ``Interactive Field Lines'' is colored with a stripe pattern.
The pattern moves along the line and its drift direction and speed convey
information of the direction and the amplitude of the vector field.

%-----------------------------------------------------------------------------------
\section{Holostage-MINI at Univ. Hyogo}
%-----------------------------------------------------------------------------------

%-----------------------------------------------------------------------------------
\subsection{Hardware}
%-----------------------------------------------------------------------------------

\begin{figure}[h]
  \begin{center}
%=================<fig>=================
   \includegraphics[width=0.4\linewidth]{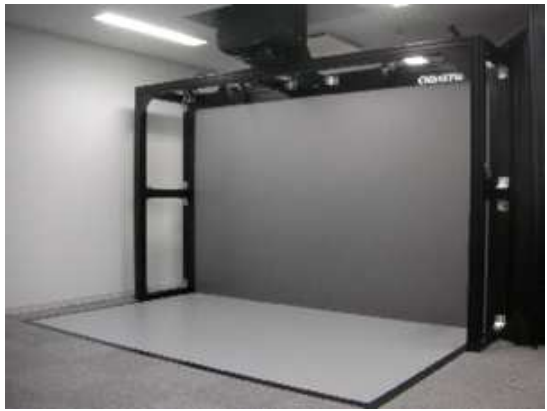}
  \caption{%
    A two-screens CAVE system installed at Univ.~Hyogo. 
    The size of the both screens is 3.2m $\times$ 2.0m.
  }
  \label{fig:121210_holomini_hyogo}
%=================</fig>=================
  \end{center}
\end{figure}

A CAVE system called Holostage-MINI,
which is a commercial brand for a small-sized CAVE system,
is installed at Univ.~Hyogo;
Fig.~\ref{fig:121210_holomini_hyogo}.
It has two screens; a wall and a floor,
with the size of 3.2m $\times$ 2.0m for both of them.
Two DLP projectors (Christie Mirage WU3), 3000 ANSI lumens, are used.
The images of the wall screen are projected from the rear.
Those of floor screen are from the ceiling.
The image resolution is 1920$\times$1200 pixels.
An optical wireless tracking system (Vicon) for the head and the hand is adopted.
Four cameras are installed on the ceiling.
Active shutter liquid crystal shutter glasses are used for the parallax stereo view.

A Linux-based computer (SGI Asterism ID412) is used for computations and graphics,
with 96 GB memory and two sets of NVIDIA Quadro FX~5800 are used as GPU.

CAVElib (ver.~3.2) is used for the basic API in all applications. 
For the tracking system, Trackd is used.

%-----------------------------------------------------------------------------------
\subsection{Applications of VFIVE}
%-----------------------------------------------------------------------------------

At Univ.~Hyogo, an important application, or its descendant, 
of VFIVE is under development.
It is called Mobile-VFIVE.

\begin{figure}[h]
  \begin{center}
%=================<fig>=================
   \includegraphics[width=0.4\linewidth]{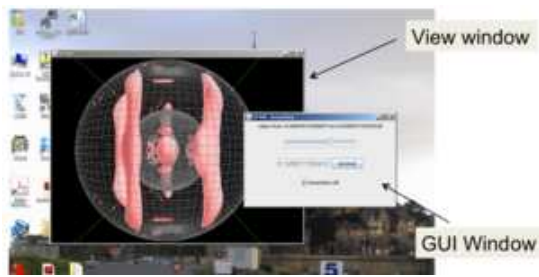}
  \caption{%
    A snapshot of Mobile-VFIVE.
    Mobile-VFIVE is a multi-platform visualization application derived from VFIVE
    for Mac, Linux, and Windows-PCs.
  }
  \label{fig:121210-mobile-vfive}
%=================</fig>=================
  \end{center}
\end{figure}

The purpose of Mobile-VFIVE is to make a common visualization
environment between CAVE and PC.
Since Mobile-VFIVE is coded in JAVA with JOGL (Java Binding for OpenGL),
in contrast that original VFIVE is coded in C++\cite{Kageyamaa},
it runs on Mac, Linux, and Windows operating systems.

A problem for power users of VFIVE was that
they could make same visualizations on their PC.
Even if a user has found an intriguing phenomenon
in his/her simulation data in the CAVE by applying 
a couple of visualization methods of VFIVE,
he/she cannot examine it in detail later on his/her PC.

The purpose of Mobile-VFIVE is to emulate VFIVE in PC.
Converting the source codes into JAVA,
most visualization methods in VFIVE are implemented.
The input data format is the same as VFIVE.
By adding a slight modification, 
one can save visualization states, such as 
applied visualization methods and their parameters,
in the VFIVE at Univ.~Hyogo.
Mobile-VFIVE realizes the visualization by
reading the state file.

Due to the hardware difference,
the user interface (UI) of Mobile-VFIVE on PC is very different from
that of VFIVE in CAVE.
The UI of Mobile-VFIVE is developed on the Swing.
When the user has found an interesting visualization
in Mobile-VFIVE, he/she can save the sate into a file.
The modified VFIVE can read the file and show the
visualization in the immersive environment of CAVE.

%-----------------------------------------------------------------------------------
\section{Holostage at Chuo Univ.}
%-----------------------------------------------------------------------------------

%-----------------------------------------------------------------------------------
\subsection{Hardware}
%-----------------------------------------------------------------------------------

\begin{figure}[h]
  \begin{center}
%=================<fig>=================
   \includegraphics[width=0.4\linewidth]{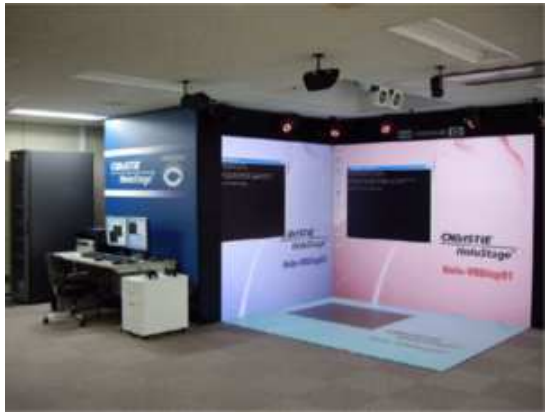}
  \caption{%
    A snapshot of Holostage at Chuo University.
  }
  \label{fig:holostage_chuo_univ}
%=================</fig>=================
  \end{center}
\end{figure}

Holostage is a commercial brand name for rectangular-box-type CAVE systems.
A Holostage is installed at Chuo University.
It is a three-screens system with two walls and a floor; see Fig.\ref{fig:holostage_chuo_univ}.
The size of the screens are W2.8m $\times$ D2.1m  $\times$ H2.1m.

An optical tracking system (Vicon) is used for the head and hand tracking.
Computer system is a Windows-based PC cluster.
CAVElib and Trackd are used as the basic APIs.

%-----------------------------------------------------------------------------------
\subsection{Applications of VFIVE}
%-----------------------------------------------------------------------------------

One of the most serious limitations of the original VFIVE is
the input data format.
An important improvement of VFIVE is applied at Chuo Univ.~to resolve this limitation.

\begin{figure}[h]
  \begin{center}
%=================<fig>=================
   \includegraphics[width=0.4\linewidth]{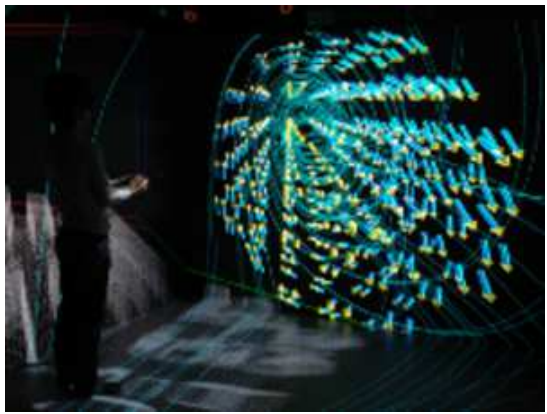}
  \caption{%
    A snapshot of a visualization application derived from VFIVE.
    It accepts unstructured cell data (UCD); in contrast that
    original VFIVE accepts only rectilinear coordinate data.
  }
  \label{fig:ucd-vfive-chuo}
%=================</fig>=================
  \end{center}
\end{figure}

The original VFIVE accepts only structured data, 
defined on the so-called rectilinear coordinates $x$-$y$-$z$,
in which grid spacing in each direction of $x$, $y$, and $z$ can be non-uniform.
There was a version of VFIVE that accepted general structured data on 
curved coordinates, such as the spherical polar coordinates\cite{Kageyama2000}.
The version is not maintained more.

Kashiyama et al.~have developed a new visualization application 
that accepts unstructured cell data (UCD)\cite{Kashiyama}.
All the basic visualization methods such as ``Particle Tracers'',
``Field Lines'', ``Local Arrows'' in the original VFIVE were 
rewritten for the UCD data. 
A snapshot of this application is shown in Fig.~\ref{fig:ucd-vfive-chuo}.

Technical challenges in the development reside in the interpolation.
Spatial interpolations are frequently called in many visualization methods in VFIVE.
Field data $\phi(x,y,z)$ at a position $(x,y,z)$,
that is specified by the wand, for example,
has to be interpolated from $\phi(x_i, y_i, z_i)$ values
where $(x_i,y_i,z_i)$, with $i=1,\ldots, N$,  are $N$ nearest grid points.

To realize the real time response with the head tracking in CAVE,
high-speed interpolations are indispensable.
The interpolation itself is just a 
linear sum $\phi(x,y,z)=\sum_i^N w_i \phi(x_i,y_i,z_i)$
with weights $w_i$ and it does not cost so much.
Delaying factor is in the search of the nearest neighbors $(x_i, y_i, z_i)$.
In the rectilinear coordinates, the search is
completed in three searches of $O(N)$;
in $x$, $y$, and $z$ directions.
On the contrary, the implementation
of fast algorithm of the grid search in unstructured cells is technically challenging.
Kashiyama et al.~have resolved this problem\cite{Kashiyama}.

%-----------------------------------------------------------------------------------
\section{BRAVE at Earth Simulator Center}
%-----------------------------------------------------------------------------------

%-----------------------------------------------------------------------------------
\subsection{Hardware}
%-----------------------------------------------------------------------------------

BRAVE is a CAVE system installed at Earth Simulator Center.
The configuration is a cubic, as the original CAVE system,
surrounded by four screens (three wall screens and a floor screen).
The side length is 3.0m.
The DLP projectors (Christie Digital Systems, MirageS+3K), 3000 ANSI lumens,
are located behind the wall screens.
Another projector is placed on the ceiling for the floor.
The image resolution is 1050$\times$1050 in each screen.
14 sets of infrared emitters are placed around the CAVE room
for the image synchronization.
An optical tracking system (Vicon MX) is used for the
head and the hand tracking.
Four cameras are installed on the ceiling.

\begin{figure}[h]
  \begin{center}
%=================<fig>=================
   \includegraphics[width=0.4\linewidth]{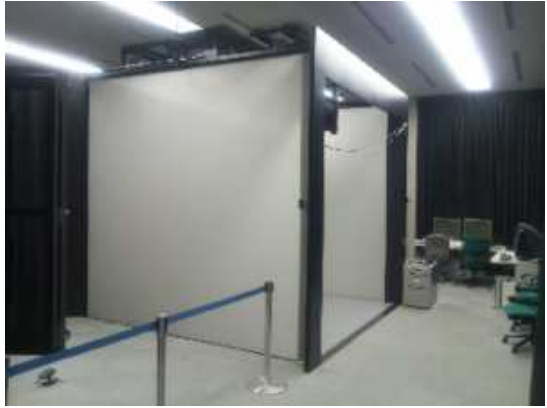}
  \caption{%
    BRAVE system installed at Earth Simulator Center.
  }
  \label{fig:121210_BRAVE}
%=================</fig>=================
  \end{center}
\end{figure}

A Linux-based computer (SGI Asterism Ultra AO532)  
with 256 GB memory
and two sets of NVIDIA Quadro PLEX 1000 Model IV is used.
Two Quadro FX~5600 are installed in each Quadro PLEX.

CAVElib (ver.~3.2) is used for the basic API in all applications
used in the BRAVE, including VFIVE.
For the tracking system, Trackd is used.

%-----------------------------------------------------------------------------------
\subsection{Applications of VFIVE}
%-----------------------------------------------------------------------------------

BRAVE system is used mainly for geoscientific data
produced by large-scale computer simulation,
such as typhoon, geodynamo, mantle convection, magnetosphere simulations.

A lot of visualization methods and functions implemented in VFIVE
were developed at BRAVE such 
as the volume rendering, animation, VTK incorporation, and Region of Interest (ROI).
See our paper\cite{kageyama1998} for details.

%-----------------------------------------------------------------------------------
\subsubsection{Geodynamo}
%-----------------------------------------------------------------------------------

One of the most actively applied visualizations of VFIVE in
the BRAVE is for geodynamo simulations.
The purpose of this simulation
is to understand the generation mechanism of the
geomagnetic field.
Large-scale
geodynamo simulations\cite{Kageyama2008,Miyagoshi2010} are performed
on Earth Simulator Supercomputer.
Since the spatial resolution of this calculation
is the highest in this kind of simulations,
the produced data size is enormous.
For the geodynamo simulation, the data loaded 
to VFIVE usually amounts to 17~GB in total (17 sets of 1~GB data).

The visualization of output data of geodynamo simulations
is much more difficult compared with a simple CFD data
with the same size.
Three-dimensional variables that should be visualized in the geodynamo
simulation are four vector fields (flow field, vorticity field,
magnetic field, and electrical current field)
and, at least, two fields (temperature and pressure).

\begin{figure}[h]
  \begin{center}
%=================<fig>=================
   \includegraphics[width=0.4\linewidth]{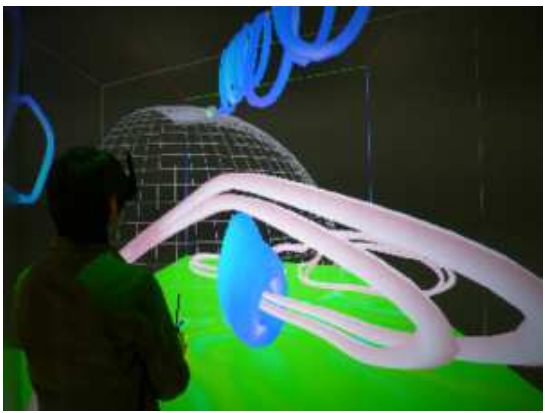}
  \caption{%
    An example of scientific discovery made possible by VFIVE.
    The lines of force of the electric field current (blue curves in the figure)
    in a geodynamo simulation data 
    exhibit an expected structure of a closed torus-like structure.
  }
  \label{fig:121207_P1000378_dynamo_coil}
%=================</fig>=================
  \end{center}
\end{figure}

Through interactive visualizations with VFIVE, 
a new structure of the electric field was found.
When the ``Field Line'' method of VFIVE was applied to fields,
they found that the magnetic field lines tend to be
straight lines (this was expected),
and electric current field lines tend to have
coil-like structure (this was also expected).
However, they were surprised that some of the electric current coils
exhibit a torus-shape in which a line covers a closed surface of a torus.
The pink curves in Fig.~\ref{fig:121207_P1000378_dynamo_coil} are
magnetic field lines, and the blue curves are electric current lines.
The blue object in near the center of the picture is the closed-torus
of the electric current line found in CAVE.
This would probably be the most important discovery by VFIVE\cite{Kageyama2008}.

%-----------------------------------------------------------------------------------
\subsubsection{User Interface using Mobile Devices}
%-----------------------------------------------------------------------------------

\begin{figure}[h]
  \begin{center}
%=================<fig>=================
   \includegraphics[width=0.5\linewidth]{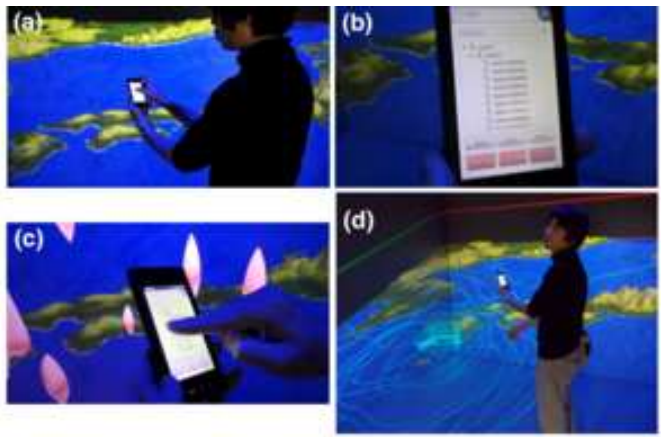}
  \caption{%
    A mobile interface applied in VFIVE visualization. 
    Visualization methods as well as parameters are 
    automatically recorded through a VFIVE visualization session.
    (a) The recorded data are shown on the device display.
    (b) The user specifies a visualization method in the list, then their internal parameters.
    (c) In this case, an isosurface level of a scalar field (cloud water density) was 
    selected in the previous step~(b) and the isosurface visualization method is 
    re-applied with the that level and shown.
    (d) In this snapshot, stream lines with a vortical structure (a typhoon) was selectively
    displayed by re-applying a proper sets of visualization parameters, that are the
    releasing positions of the stream lines.
  }
  \label{fig:121207_kawahara_combined}
%=================</fig>=================
  \end{center}
\end{figure}

Recently, 
researchers at Earth Simulator Center\cite{Kawahara2012}
have developed a new user interface for CAVE systems.
The purpose of this interface is to control
VFIVE with mobile devices such as iPad, iPhone, and Android 
devices. See Fig.~\ref{fig:121207_kawahara_combined}.

In this version of VFIVE,
a user brings a mobile device while the VFIVE application 
is running in CAVE.
A newly developed web application runs on the device.
A list of visualization methods is shown on the device display.
It tells visualization methods that have been used in the VFIVE session so far.
The user can select one of the methods
though the usual touch-display interface of the mobile device.
The display then shows a detailed list of visualization parameters
with radio buttons and check boxes.
They are parameters that have been applied in the session.

When buttons or boxes are selected,
the information is sent to the running VFIVE through the wireless network.
VFIVE re-apply the selected visualization operations with specified parameters.

It is common in VFIVE visualization that
its user applies a lot of visualization methods
with many different visualization parameters
in each method.
In this improvement of VFIVE,
a sequence of visualization operations applied 
by the user is automatically stored.
The user can make VFIVE ``remember'' the best visualizations
from the history through the mobile device.

%-----------------------------------------------------------------------------------
\section{CompleXcope at NIFS}
%-----------------------------------------------------------------------------------

%-----------------------------------------------------------------------------------
\subsection{Hardware}
%-----------------------------------------------------------------------------------

\begin{figure}[h]
  \begin{center}
%=================<fig>=================
   \includegraphics[width=0.40\linewidth]{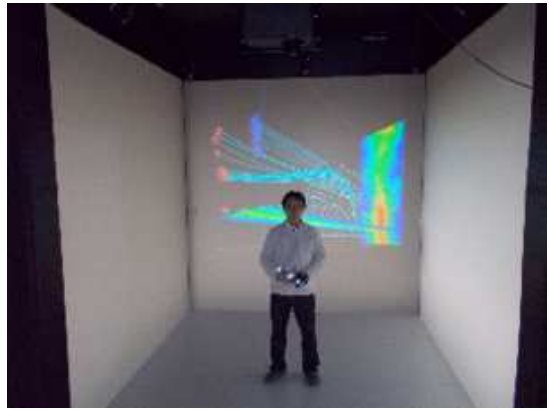}
  \caption{%
    Overview of CompleXcope at National Institute for Fusion Science.
    A cubic CAVE with 3m sides.
    This CAVE system is used for nuclear fusion science and plasma physics.
  }
  \label{fig:121207_Xcope_overview}
%=================</fig>=================
  \end{center}
\end{figure}

CompleXcope (Fig.~\ref{fig:121207_Xcope_overview}) is a CAVE system installed at
National Institute for Fusion Science (NIFS).
The configuration is a cube with three walls and floor.
The side length is 3m.

An optical system (ARTtrack) is used for 
the head and hand tracking.
Five cameras are installed on the ceiling.
Three-dimensional sound system is also installed.
Four DLT projectors (Christie Mirage S+3K) are used.
The image resolution is 1050~x~1050 for each screen.

Two sets of computer systems are connected.
One is Linux-based system (SGI Asterism) with 32 GB memory
and Quadro FX~5800 GPU.
Another is Windows-based PC cluster system.
Two PCs (HP Z820) are connected.
Each PC has 64 GB memory and 
two sets of NVIDIA Quadro 5000.
For the basic APIs, CAVElib (ver.~3.2.1) and Trackd are used.

Several application programs, including VFIVE, are used 
in CompleXcope. 
Among them,
an application program was developed 
to visualize a high temperature plasma equilibrium state 
in LHD (Large Helical Device) of National Institute for Fusion Science{\cite{Ohtani2012}.

%-----------------------------------------------------------------------------------
\subsection{Applications of VFIVE}
%-----------------------------------------------------------------------------------

Magnetic reconnection\cite{Biskamp2005} is 
one of the most important processes in plasma physics observed in
laboratory experiments, magnetospheric substorms,
solar flares, and so on.
The magnetic energy is converted into
plasma particle kinetic energy through the magnetic reconnection.
When two magnetic field lines with an angle,
a reconnection takes place at so-called X-point.
The plasma particles, which are composed of ions and electrons,
exhibit highly complex motions around the X-point.
It is important to examine these particle trajectories.

\begin{figure}[h]
  \begin{center}
%=================<fig>=================
   \includegraphics[width=0.40\linewidth]{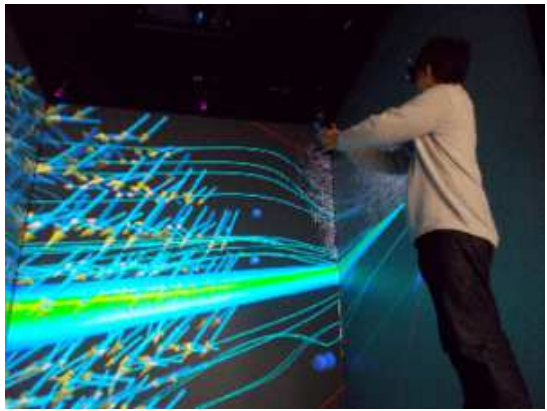}
  \caption{%
    Visualization of plasma particle simulation data with VFIVE.
  }
  \label{fig:121207_Xcope_vfive}
%=================</fig>=================
  \end{center}
\end{figure}

A new function has been added on the VFIVE at CompleXcope.
The purpose is to grasp the three-dimensional complex orbits of the particles.
Particle motions under electromagnetic fields are calculated in
real time\cite{Ohtani2008}. 
This is a kind of test particle simulation incorporated 
into VR visualization.
First, VFIVE reads a snapshot data of electromagnetic fields
obtained by a plasma simulation.
As in other VFIVE applications, user can analyze the
electromagnetic fields through various kinds of visualization
methods implemented in the standard VFIVE.
After selecting a newly added menu panel of VFIVE,
the user specifies an initial position of a test particle, an ion or electron, 
with the wand.
This particle is released 
with an initial velocity vector, which is a function of position
calculated from the simulation data.
The equation of motion of the particle is integrated in
real time by a numerical integrator with the electromagnetic 
forces of the loaded data.
One can observe the test particle motion in 3D in the CompleXcope's
VR space.
Refer\cite{Ohtani2008} for details of the implementation and 
scientific results with this new visualization method.

In the above visualization, plasma particle
motions are analyzed under fixed electromagnetic fields,
assuming no temporal variations.
Recently, Ohno and Ohtani\cite{Ohno2012} have improved this method
for dynamic analysis with temporal variations.
A set of electromagnetic fields at 
tens of different time steps,
which were produced by a supercomputer simulation,
was transferred to the local HDD of CompleXcope.
VFIVE reads the sequence of fields one after another.
The user can apply any visualization method
implemented in VFIVE
to a specified time step data.
Selected visualization method, for example the particle tracer, 
is applied not only to the start time step, but also every step after that.
This makes an animated visualization of particle tracer 
under time varying fields.
After the first cycle of this method,
calculated visualization objects, such as polygons for
the particle obits, are stored.
The stored data can be replayed later, making it possible
to review quickly the whole animated visualization.

%-----------------------------------------------------------------------------------
\section{Conclusion}
%-----------------------------------------------------------------------------------

In this paper, we have summarized both the developments
and scientific achievements made possible by VFIVE
in several CAVE systems in Japan.

In contrast to commercially available visualization applications
for CAVEs, our visualization application VFIVE 
has only a basic set of visualization methods.
The compactness of the application is a natural consequence
of its history.
It was first developed in late 1990s by simulation researchers who were
eager to analyze their own simulation data in a CAVE.
Implemented visualization methods and user interface
were, therefore, minimum.
Since then, VFIVE has been improved under
constant feedbacks from a broader range of users 
where CAVE systems are installed.
The compactness of the program
has helped the users and developers adding new functions.

Through continued developments and improvements,
VFIVE has become a practical tool
for scientific visualization in CAVE systems
and fruitful scientific results have been attained by this application.

%
%========================================================
\section*{Acknowledgement}
%========================================================
%
This work was supported by
JSPS KAKENHI Grant Numbers 19740346, 23340128
and Takahashi Industrial and Economic Research Foundation.
It was also performed with the support and under the 
auspices of the National Institute for Fusion Science (NIFS) Collaborative Research
Program (NIFS09KDBN004).

%\bibliographystyle{unsrt}
%\bibliography{121201}

\end{document}